\documentclass[10pt,twocolumn,twoside]{IEEEtran}
\usepackage{amssymb,amsmath}  
\usepackage{graphicx}        
\usepackage{float}  		
\usepackage{url}
\usepackage{float} 
\usepackage{psfrag} 
\usepackage{url}
\usepackage{subfig}

\usepackage{chemarr}
\usepackage{fancybox}
\usepackage{multicol}
\usepackage{wrapfig}
\usepackage{stfloats}
\usepackage{mathtools, cuted}
\usepackage{url}
\usepackage{cite}
\usepackage{epstopdf}
\usepackage{xcolor}
\usepackage{tabularx}
\usepackage{booktabs} 
\usepackage[margin={1.60cm,1.60cm}]{geometry}
\usepackage{soul}
\begin{document}

\title{Introduction to the Special Issue on Approaches to Control Biological and Biologically Inspired Networks}
\author{R\'eka Albert, John Baillieul, and Adilson E. Motter\\ {IEEE Trans. Control Netw. Syst. 5(2), 690-693 (2018)}
\thanks{
R.\ Albert is with the Departments of Physics and Biology, Pennsylvania State University, University Park, Pennsylvania 16802, USA. 

J.\ Baillieul is with the Department of Mechanical Engineering and the Division of Systems Engineering, Boston University, Boston, Massachusetts 02215, USA.

A.\ E.\ Motter is with the Department of Physics and Astronomy and the Northwestern Institute on Complex Systems, Northwestern University, Evanston, Illinois 60208, USA. Email: motter@northwestern.edu.}
}

\maketitle

\begin{abstract}
The emerging field at the intersection of quantitative biology, network modeling, and control theory has enjoyed significant progress in recent years. This Special Issue brings together a selection of papers on complementary approaches to observe, identify, and control biological and biologically inspired networks. These approaches advance the state of the art in the field by addressing challenges common to many such networks, including high dimensionality, strong nonlinearity, uncertainty, and limited opportunities for observation and intervention. Because these challenges are not unique to biological systems, 
it is expected that many of the results presented in these contributions will also find applications in other domains, including physical, social, and technological networks.

\end{abstract}


\IEEEpeerreviewmaketitle

\IEEEPARstart{A}{}lthough networks of various forms have existed throughout history, it is only comparatively recently that recognizable network technology sectors have emerged as important components of the collective human enterprise.  After their introduction, power grids and telephone networks underwent 
 rapid expansion and increasing interconnection throughout the first half of the twentieth century.  Even though these networks shared certain characteristics, their dissimilarities were such that common principles of design and analysis were rarely proposed.  Unifying approaches to understanding networks across a variety of technological domains began to emerge in the final decades of the century, after a number of industrial control technologies were introduced---perhaps the most notable being the controller area network 
(CAN) bus, which is the message-based 
bus standard for networking microcontrollers in automobiles introduced by Bosch GmbH in 1986.  An early reference on this and other pioneering industrial control networks of the 1980s and 1990s is \cite{Bushnell}.  Since the turn of the 
century, there has been an
explosion of interest in the basic principles underlying the technology of network control systems.  While much
of this has been driven by the rapid penetration of wireless communication technologies (see, e.g., \cite{TAC2004, Hristu, Proceedings2007}),
this interest was also accompanied  by an increasing focus on various types of networks within the biological domain (for an early reference, see \cite{CSM2004}).

Over the past decade, the scientific and technical literature dealing with control of biological networks has grown  
rapidly in breadth and depth. 
This progress has been determined by 1) the recognition that many biological systems are best described as networks of interacting components, 2) the increased availability of data on the structure and dynamics of biological networks, and 3) the advances on experimental approaches to sense and 
 actuate individual components of such systems. Because biological networks can be used to describe many processes, their control is broadly significant both to reveal naturally evolved control mechanisms underlying the functioning of biological systems and to develop human-designed control interventions to recover lost function, mitigate failures, and repurpose the system. Application areas have included molecular and cell biology, neuroscience, and ecology as well as biologically inspired engineering applications such as collective formations involving moving sensors.

In neuronal networks, for example, it is of interest to understand and influence the collective dynamics of neurons, as well as investigate their relation to the sensory system and to motor control and other outputs. In intracellular networks, understanding the workings of the regulatory system---a control system par excellence---has 
much to contribute to the identification of therapeutic interventions and the development of synthetic biology. In ecological networks, network-based measures to correct imbalances have been proposed as useful ecosystem-management tools to help prevent species extinctions. Decentralized control of multi-agent systems, on the other hand, is an application area of network control that speaks to numerous natural as well as engineered systems.

In biology, as in other fields, the central role of control is to induce desired behaviors and prevent undesired ones. There are, however, salient properties that set biological networks of interest apart from the  
typical  low-dimensional engineered systems traditionally considered in theoretical studies. Such properties may include: (i) limited ability to measure the dynamical state of the system, (ii) presence of parameter uncertainty or lack of predictive mathematical models, (iii) high dimensionality of the state and parameter spaces, (iv) strong nonlinearity and multi-stability of the underlying dynamics, (v) strict constraints on implementable control interventions, (vi) decentralized evolution and operation, and (vii) limited opportunities for the implementation of feedback. These properties can make it difficult to recognize control mechanisms that are both effective and efficient in biological networks.

The resulting challenges, which often appear in tandem, are theoretical and computational in nature. For example, even when experimental techniques exist to actuate the individual components of a biological network---such as in the control of the expression of individual genes in a regulatory network---the systematic design of a network-wide intervention to achieve a predefined objective cannot be addressed by purely experimental means due to a combinatory explosion in the number of possibilities. The systematic solution of such problems calls for scalable theoretical control approaches that can in addition adequately represent the experimental conditions. Concrete progress has recently been made toward addressing these challenges in the contexts of control (e.g., \cite{cornelius2013realistic,zanudo17structural}) as well as observability (e.g., \cite{liu2013observability,whalen2015observability}) of complex biological networks, and this Issue brings together important new additions to this rapidly growing body of literature. 

The contributions in this Special Issue cover a broad range of timely problems relevant to the interface between biological systems, network dynamics, and control theory.   Four articles present on progress toward improving the identifiability 
and observability of network systems, with particular emphasis on biological networks: 

\begin{itemize} 

\item
Haber, Molnar, and Motter \cite{haber} introduce a method for optimally selecting sensor nodes and reconstructing the state from limited information in nonlinear network systems.
The method, which is devised as an optimization-based approach, naturally accounts for modeling uncertainties and can be applied to very large networks.
The effectiveness of the approach is demonstrated in chemical, signaling, and regulatory networks, where it identifies the key  nodes from which the full network state  can be determined.

\item
Tzoumas, Xue, Pequito, Bogdan, and Pappas \cite{tzoumas} consider the problem of sensor selection in biologically motivated discrete-time linear fractional-order systems,
which are   systems  characterized by involving a fractional-order difference operator (the discrete counterpart of a fractional derivative). 
The approach aims at identifying the  minimum number of state variables that  need to be measured to monitor the evolution of the entire network. 
The biological significance of the approach in the presence of noise and  modeling errors is illustrated using electroencephalogram data.

\item
Nozari,  Zhao, and Cort\'es \cite{nozari} address the problem of network identification when only a fraction of the nodes can be sensed and actuated in  systems with 
  linear time-invariant dynamics. The approach presented can determine the transfer function and network structure of the set of accessible nodes,  by means of an 
  auto-regressive model. Application of the approach to time-series of electroencephalogram data recorded 
  from the human brain illustrates its potential when the number and the state of the remaining nodes in the network is unknown.
 
 \item
Pan,  Yuan,  Ljung,  Gon\c{c}alves, and Stan \cite{pan} introduce a method to identify the parameters of nonlinear systems using heterogeneous data. The method adopts 
a sparse Bayesian formulation to infer the sparsest model that can explain the  available data. The method is
well suited for biochemical networks, where it can determine both the reaction dynamics and kinetic constants from data generated
through a potentially large number of different experiments.

\end{itemize}

Three articles consider networks described by Boolean dynamics, which are often used to model regulatory, 
signaling, and gene networks  in general, with special attention given to the control of  such networks toward states or attractors of interest:

\begin{itemize} 

 \item
Clark,  Lee, Alomair, Bushnell, and Poovendran \cite{clark} present a computational approach to identify the minimal set of control nodes to direct a Boolean  network to a desired attractor.
Specifically, the authors establish a sufficient condition for convergence to an attractor, which allows recasting the problem as one that  can be solved 
using integer linear programming. Application of the approach to various regulatory and signaling networks shows 
that it can find use 
in the design of strategies for 
cell reprogramming and therapeutic interventions. 

 \item
Imani and Braga-Neto \cite{imani} focus on the development of an approximate control approach for gene regulatory networks with uncertainty 
both in the measurements and interventions.
The approach is based on modeling the networks as partially-observed Boolean dynamical systems and applying a combination of reinforcement learning and 
Gaussian process techniques.  Numerical simulations show promise for large  regulatory networks whose state is determined from noisy gene expression data.

 \item
Gao, Chen, and Ba\c{s}ar \cite{gao} establish necessary and sufficient conditions for conjunctive Boolean networks to be controllable toward specific orbits or states.
Such networks, which can be used in the modeling of genetic networks, are characterized by having only  the {\sc and} logic operation. Their analysis leads 
to explicit control laws, which involve control of the cycles in the network, thereby addressing the algorithmic problem of prescribing the control inputs to steer the network.

\end{itemize}

Finally, partially motivated by the potential to inform the design of synthetic systems, the other three articles of the Issue address the relationship between the network structure and the dynamic 
repertoire of networked systems:

\begin{itemize} 

 \item
Blanchini, Samaniego, Franco, and Giordano \cite{blanchini} consider networks that can be decomposed into subsystems exhibiting monotonic step response (MSR), which is characterized by monotonic 
behavior of the output function at the equilibria points of the system.  By examining the positive and negative cycles of an aggregated graph, whose nodes are 
the MSR subsystems, they are able to classify possible multi-stationary and oscillatory dynamical behaviors of the system. These results  are relevant for the
modeling of  biological networks and the design of bio-inspired systems that can be approximated as aggregates of MSR subsystems.

 \item
Gray, Franci,  Srivastava, and Leonard \cite{gray} present an agent-based model for collective decision-making in multi-agent networks. The model, which is inspired 
by the high-performing decision-making dynamics implemented by honeybees, establishes a direct relation between mechanisms 
identified in collective animal behavior and bio-inspired control designs for choosing among alternatives in engineered multi-agent
network systems.

 \item
Steel and Papachristodoulou \cite{steel} identify constraints that can be used to design synthetic biological networks able to achieve two fundamental properties common
to many natural biological systems: homeostasis and adaptation. The identified constraints can orient selection of network topologies that  both
exhibit the desired behavior and are easily implementable.  The results have potential to foster experimental construction of increasingly complex
systems in the area of synthetic biology.

\end{itemize}

The theme of this Special Issue is thus centered on quantitative approaches to observe and control the behavior of biological networks and to model naturally evolved network-control mechanisms in living 
systems. The main focus is on theoretical and computational approaches that are constrained by data, applied to realistic models, or otherwise account for important salient features of real biological or 
biologically inspired networks. Taken together, these approaches have the potential to help address numerous pressing research questions, including the  development of rational techniques for cell 
reprogramming and transdifferentiation, the design of interventions to mitigate disease states, the control of neural and brain networks, and the formulation 
of  improved strategies for 
ecosystem management. They may also lead to more systematic methods for  system identification, model construction, and state observation  as well as a
 better 
understanding of the  interplay between stability, control, and robustness in biological networks. Finally, by helping to reveal the mechanisms underlying the dynamics of natural 
biological networks, these approaches have a unique potential to foster the development of biologically-inspired network designs in diverse application areas.

\section*{Acknowledgement}

The Guest Editors thank the Mathematical Biosciences Institute (MBI) at The Ohio State University, where they had the opportunity to participate in the
Emphasis Program ``Dynamics of Biologically Inspired Networks" as organizers of the highly successful  workshop ``Control and Observability of Network Dynamics" on 
April 11-16, 2016. That workshop
served as inspiration and reference for this Special Issue, and indeed several open questions discussed
at that meeting are addressed in contributions included herein.

\begin{IEEEbiography}
[{\includegraphics[width=1in,height=1.25in,clip,keepaspectratio]{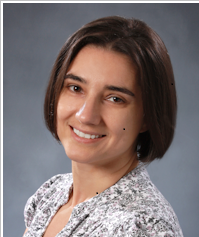}}]
{R\'eka Albert}
is a Distinguished Professor of Physics at The Pennsylvania State University, with adjunct appointments in the Department of Biology and the Huck Institutes of the Life Sciences. She received her Ph.D. in Physics from the University of Notre Dame in 2001, working with Prof. Albert-L\'aszl\'o Barab\'asi. She then did postdoctoral research in mathematical biology at the University of Minnesota, working with Prof. Hans G. Othmer.  Prof. Albert's research is focused on predictive modeling of the dynamics of biological networks at multiple levels of organization.  She is a Fellow of the American Physical Society (APS) and an External Member of the Hungarian Academy of Sciences (MTA). 
\end{IEEEbiography}

\vspace{-1cm}

\begin{IEEEbiography}
[{\includegraphics[width=1in,height=1.25in,clip,keepaspectratio]{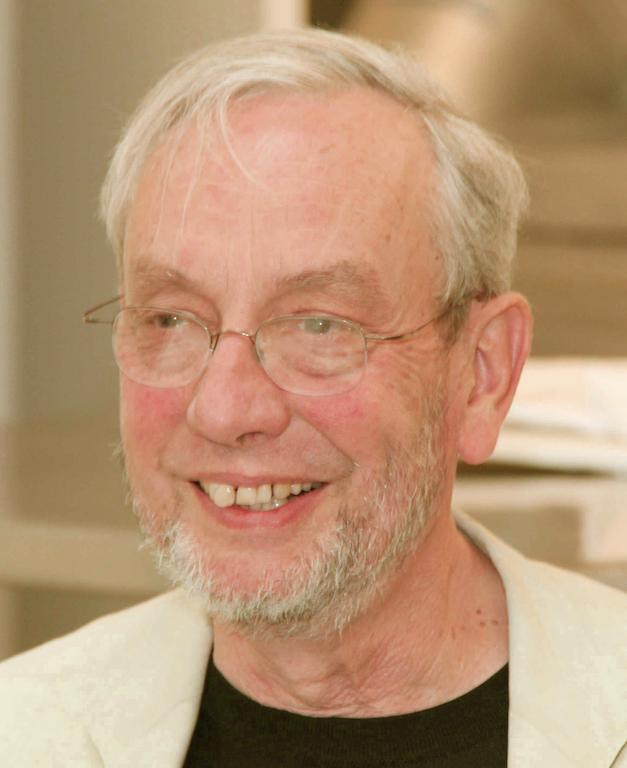}}]
{John Baillieul}
(Life Fellow, IEEE) is a Distinguished Professor of Engineering at Boston University. 
He received his Ph.D. in Applied Mathematics from Harvard University in 1975.
He has published research on robotics, the control of mechanical systems, and more generally in mathematical system theory. His early work dealt with connections between optimal control theory and what came to be called sub-Riemannian geometry. Other early work treated the control of nonlinear systems modeled by homogeneous polynomial differential equations. 
His main controllability theorem applied the concept of finiteness embodied in the Hilbert basis theorem to develop a controllability condition that could be verified by checking the rank of an explicit finite-dimensional operator. 
He also worked on the qualitative theory of electric energy systems, and over the past decade
 he has turned his attention to the study of information-based control. 
Prof. Baillieul is a Fellow of the International Federation of Automatic Control (IFAC) and of the Society for Industrial and Applied Mathematics (SIAM).
\end{IEEEbiography}

\vspace{-1cm}

\begin{IEEEbiography}
[{\includegraphics[width=1in,height=1.25in,clip,keepaspectratio]{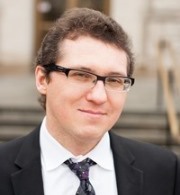}}]{Adilson E. Motter}
is the Charles Morrison Professor of Physics and a member of the Northwestern Institute on Complex Systems (NICO) at Northwestern University. He received his Ph.D. degree in 2002 from UNICAMP and his postdoctoral training from the Max Planck Institute for the Physics of Complex Systems and Los Alamos National Laboratory. His research is focused on the dynamical behavior of complex systems and networks. 
He has contributed to the control and observability of nonlinear network systems, the  modeling of cascading failures and synchronization phenomena in networks, 
and the discovery of synthetic rescues in metabolic networks, longitudinal negative compressibility in mechanical metamaterials, and the phenomenon of asymmetry-induced symmetry 
(in which the stability of symmetric states require system asymmetry).
Prof. Motter is a Fellow of the American Physical Society (APS) and of the American Association for the Advancement of Science (AAAS).
\end{IEEEbiography}

\vfill

\end{document}